\documentclass[conference]{IEEEtran}
\IEEEoverridecommandlockouts
\usepackage{cite}
\usepackage{amsmath,amssymb,amsfonts}
\usepackage{algorithmic}
\usepackage{graphicx}
\usepackage{textcomp}
\usepackage{xcolor}
\usepackage{multirow}
\def\BibTeX{{\rm B\kern-.05em{\sc i\kern-.025em b}\kern-.08em
    T\kern-.1667em\lower.7ex\hbox{E}\kern-.125emX}}
\begin{document}

\title{Securing the Classification of COVID-19 in Chest X-ray Images: A Privacy-Preserving Deep Learning Approach}

\author{Wadii Boulila\textsuperscript{1,2}, 
        Adel Ammar\textsuperscript{1},
        Bilel~Benjdira\textsuperscript{1}, 
        Anis~Koubaa\textsuperscript{1} \\
$^1$RIOTU Lab, Prince Sultan University, Riyadh, Saudi Arabia\\
$^2$RIADI Laboratory, National School of Computer Sciences, University of Manouba, Manouba, Tunisia
}
\maketitle

\begin{abstract}
Deep learning (DL) is being increasingly utilized in healthcare-related fields due to its outstanding efficiency. However, we have to keep the individual health data used by DL models private and secure. Protecting data and preserving the privacy of individuals has become an increasingly prevalent issue. The gap between the DL and privacy communities must be bridged. In this paper, we propose privacy-preserving deep learning (PPDL)-based approach to secure the classification of Chest X-ray images. This study aims to use Chest X-ray images to their fullest potential without compromising the privacy of the data that it contains. The proposed approach is based on two steps: encrypting the dataset using partially homomorphic encryption and training/testing the DL algorithm over the encrypted images. Experimental results on the COVID-19 Radiography database show that the MobileNetV2 model achieves an accuracy of 94.2\% over the plain data and 93.3\% over the encrypted data.
\end{abstract}


\section{Introduction}

The protection of human private data is a persistent challenge. This matter has always been taken seriously in computer science. However, due to the efficiency of deep learning (DL) algorithms, many have used them without any care for the protection of private data. All have been pleased by improving the accuracy of learning and delaying any privacy concerns.\par
A small interest in this topic has been raised after the deployment of remote DL models over the internet. In fact, a variety of cloud platforms were designed using DL models for both the training and inference phases. The service offered by these platforms was coined later as Deep Learning as a Service (DLaaS) \cite{latif2021deep}. Hence, private data are transferred over the internet and may also be stored on servers with high vulnerability for leakage. The leakage can be made in an intentional or unintentional way. It can be leaked intentionally through hacking, piracy, or social engineering. It can also be leaked unintentionally by the user himself or the service provider employees. Although being unexpected, the employees are behind 43\% of the whole data leakage cases, as affirmed by Intel Security \cite{mireshghallah2020privacy}. Around half of these cases were made unintentionally. \par
The first event that triggered global interest in this problem is the Facebook data privacy scandal \cite{liu2020machine}. People have discovered that Facebook has sold the personal data of 80 million users to the British firm Cambridge Analytica to train DL algorithms to make political advertising \cite{facebook_cnbc}. People have sensed that their private data are threatened and should be protected. DL algorithms should be accompanied during the training and inference by privacy-preserving techniques that prevent any identification of personal data. \par
The proposed work in this paper focuses on this topic. We designed a data privacy technique to keep private data protected during DL training and inference. The data preserving technique presented in this paper is designed for preserving Chest X-Ray images used to distinguish COVID-19 cases from normal, lung opacity, and pneumonia cases. The DL model does not have access to any identifiable information. It takes only as input encrypted images. The encryption of the Chest X-Ray is done from the user side before feeding them to the DL model. This encryption is homomorphic, which means that the final classification result does not need any decryption of the image. The DL model will only work with these encrypted data without requiring any personally identifiable information. We think this feature is essential, especially in the current COVID-19 pandemic. In fact, several research works are conducted to show the importance of using Chest X-Ray images to detect COVID-19 cases \cite {taylor2020review,atitallah2021randomly, ben2021fusion,rehman2021contactless,guefrechi2021deep,marwa2021}. Moreover, many people explicitly require that the body details shown in these kinds of images must be kept private and secure \cite{boulila2021noninvasive}. The technique presented in this paper fulfills their requirement at a very low cost (around a 1\% decrease in accuracy).  \par
The remainder of this paper is ordered as follows: in Section II, the background section, we will discuss the research works that targeted the topics of Privacy-Preservation Deep Learning (PPDL) and homomorphic encryption. Later, in Section III, we will introduce our PPDL approach for protecting COVID-19 data during the classification task. Then, in Section IV, we will describe the experimental results that defend the utility of the proposed approach. Finally, in Section V, we conclude our work and raise the main limitations that should be targeted in the next research works.  


\section{Background}
\subsection{Privacy-Preserving Deep Learning}
Data privacy is an important issue for training and testing DL models, especially in the case of training and inferring sensitive data (e.g. health records, financial details, location logs, and satellite images) \cite{boulila2021novel,jemmali2022equity}. Many PPDL techniques focused on allowing multiple input parties to train and test DL models without revealing their private data in its original form. These techniques can be subdivided into three groups: cryptographic, perturbation, secure enclaves, and hybrid techniques \cite{liu2020survey}. Cryptographic methods are used to train and test DL techniques on encrypted data \cite{alkhelaiwi2021efficient}. This category of methods includes Homomorphic Encryption (HE), Secret Sharing (SS), Secure Multi-Party Computation (SMPC), and Garbled Circuit (GC). The perturbation methods aim to alter data values to maintain individual record confidentiality \cite{chamikara2018efficient}. This category of methods includes Differential Privacy (DP) \cite{dwork2014algorithmic,papernot2016semi,bu2020deep} and Dimensionality Reduction (DP) \cite{chanyaswad2017compressive}. In secure enclaves-based approaches \cite{tramer2018slalom} \cite{ohrimenko2016oblivious}, both the prediction model and the data are separately sent by the client and the server to a trusted, secure enclave environment for execution. On the other hand, hybrid methods aim at combining more than one PPDL technique to improve the privacy of data \cite{kumar2020cryptflow,truex2019hybrid,juvekar2018gazelle,chase2017private}. A recent survey on privacy-preserving deep learning approaches can be found in \cite{tanuwidjaja2020privacy}.

Nevertheless, many of these solutions are not efficient on complex data, but only on simple classification tasks \cite{tanuwidjaja2020privacy}, such as MNIST or CIFAR-10. Besides, they often add an important computational cost and communication overhead. Moreover, there is always a trade-off between privacy and model accuracy, due to the use of approximated activation functions.

\begin{figure*}[!h]
  \centering
  \includegraphics[width=\textwidth]{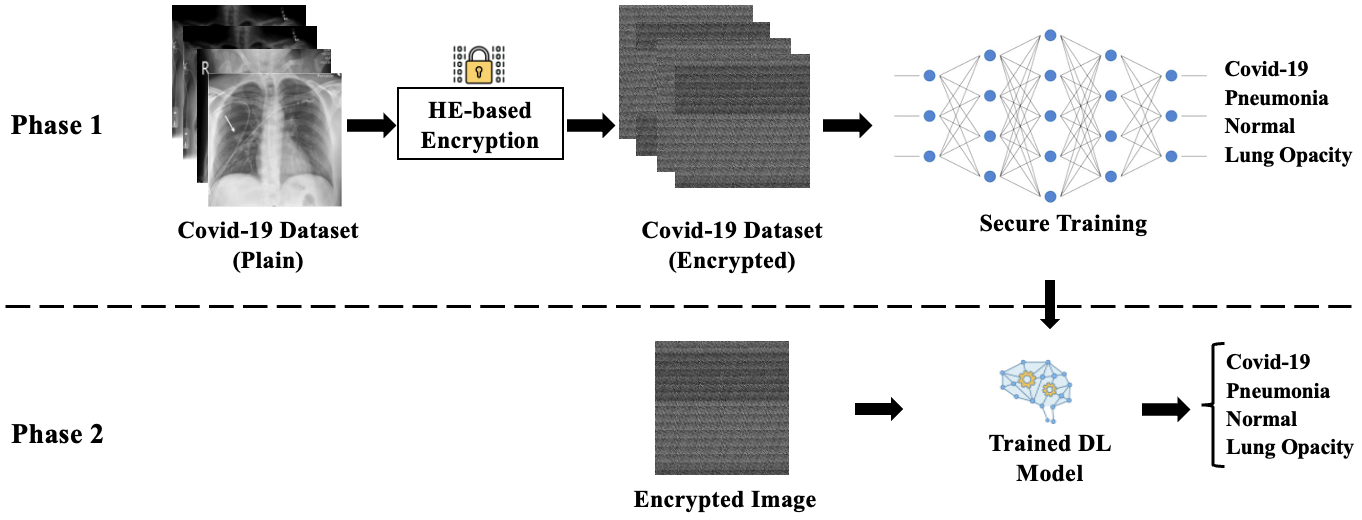}
  \caption{Proposed approach.}
  \label{fig:propsed_approch}
\end{figure*}

\subsection{Homomorphic encryption}

Most existing encryption algorithms do not allow working on data unless decrypted. However, decrypting the data undermines privacy requirements.  Once the data are encrypted by someone, they must first be decrypted before processing, which makes them vulnerable to unauthorized access.\\
HE removes the need for decrypting the data before usage. In other words, data integrity and privacy are protected while processing the data \cite{parmar2014survey,armknecht2015guide}. HE is a cryptographic technique that has the ability to allow DL techniques to run over encrypted data without losing context. It eliminates the tradeoff between data usability and data privacy and ensures that data remains secure in untrusted environments.\\
In the case of DL, the algorithm can be trained and tested over encrypted data. If the DL algorithm reaches a good prediction accuracy, it can be deployed, and therefore, in real cases, it will provide a decision over the encrypted data. Using a unique secret key, the user can decrypt the obtained data. Thus, the data’s privacy and security are maintained. The HE techniques can be divided mainly into three subcategories, namely Partially Homomorphic Encryption (PHE), Somewhat Homomorphic Encryption (SWHE), and Fully Homomorphic Encryption (FHE). PHE enables only one form of mathematical operation on the encrypted data. SHE enables all addition and multiplication operations with only a limited range on the encrypted data. FHE enables various types of assessment operations on encrypted data with an unbounded range. PHE schemes are in general more efficient than SHE and FHE, mainly because they are homomorphic with regard to only one type of operation (addition or multiplication). SHE is more general than PHE since it supports more operations; however, it can perform them on only a limited range. The main drawback of FHE is its slow computation speed.

\section{Proposed Approach}
This section describes the proposed approach for PPDL in the case of Covid-19 classification. The proposed workflow is based on two phases as depicted in Figure \ref{fig:propsed_approch}. The first phase aims to encrypt the COVID-19 dataset using the Paillier method. Then, the encrypted dataset is fed to the DL algorithm for securing training. Phase 2 allows determining the class of encrypted input images based on the trained DL model.

\subsection{HE-based Encryption}
In this paper, the Paillier method is used to ensure the encryption of images. Paillier encryption is a PHE satisfying additive homomorphism. In the following, we describe the main concepts used in this study, which are key generation, encryption, and homomorphic addition.
\begin{itemize}
    \item  \textbf{Key generation:} The public  key is ($n$,$g$) and the private key is ($\lambda$,$\mu$ ) computed as follows:
    \begin{enumerate}
        \item Compute two random and independent large-prime numbers \textit{p} and \textit{q}, where $\gcd(pq,(p-1)(q-1))=1$
        \item Compute \textit{n} and $\lambda$, where $n=pq$ and $\lambda =\operatorname {lcm}(p-1,q-1)$. lcm denotes the least common multiple.
        \item Select a random integer \textit{g}. 
        \item Compute \textit{n}, where $\mu =(L(g^{\lambda }\bmod$ $n^{2}))^{{-1}}{\bmod}$ $n$ and $L(x)={\frac {x-1}{n}}$.
    \end{enumerate}
    \item \textbf{Encryption:} Compute the ciphertext $c$, where $c=g^{m}\cdot r^{n}{\bmod}$ $n^{2}$ and $r$ is a random number ($0<r<n$ and $gcd(r,n)=1$)
    \item \textbf{Homomorphic addition:} The product of two ciphertexts will be decrypted to the sum of their corresponding plaintexts:
$D(E(m_{1},r_{1})\cdot E(m_{2},r_{2}){\bmod  n}^{2})=(m_{1}+m_{2}){\bmod}$ $n$, and the product of a ciphertext with a plaintext will be decrypted to the sum of the corresponding plaintexts:
$D(E(m_{1},r_{1})\cdot g^{{m_{2}}}{\bmod  n}^{2})=(m_{1}+m_{2}){\bmod}$ $n$.
\end{itemize}

\subsection{Secure Training}
\begin{figure}
  \centering
  \includegraphics[width=9cm]{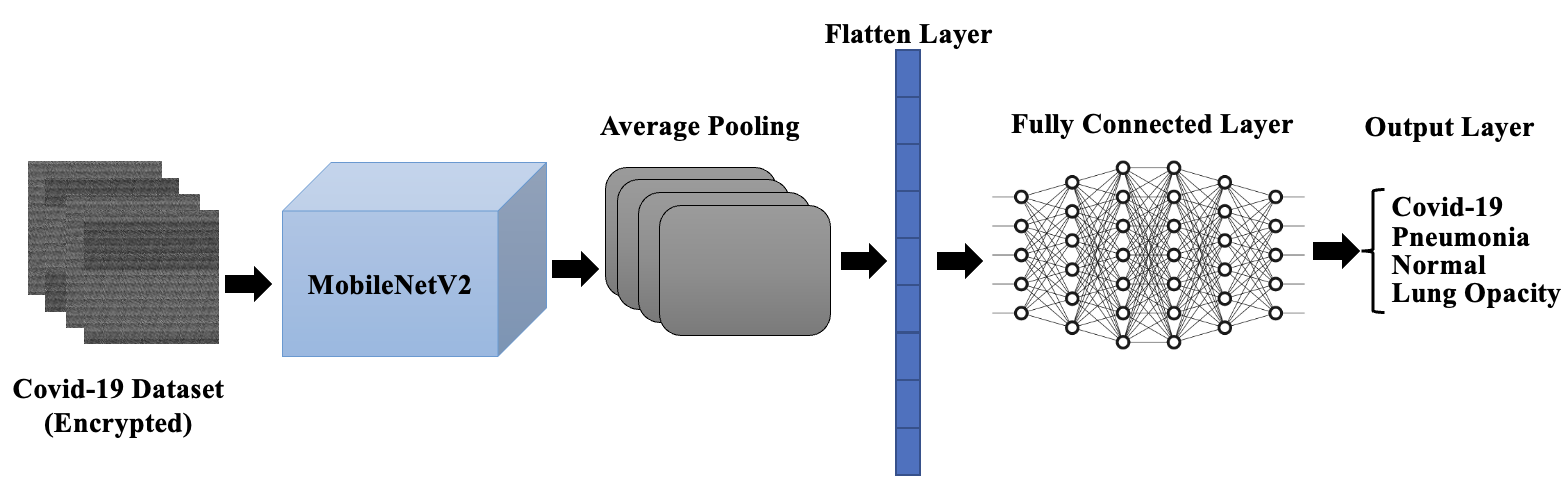}
  \caption{Architecture of the secure training algorithm.}
  \label{fig:secureTraining}
\end{figure}

The goal of the secure training is to run the DL algorithm on the encrypted dataset and be able to achieve good performance on identifying the different classes.\\
The purpose of the presented work is to evaluate the performance of DL algorithms over Paillier-encrypted images. The main challenge is to find a trade-off between the accuracy of the DL algorithm in identifying COVID-19 classes and the security of images. To assess the first point, a good alternative will be to check the performance of the selected DL algorithms on plain and encrypted data. In case of a small difference between the accuracy of the DL algorithm on plain and encrypted data, we can assert that it performs well and can be adopted in real cases. Otherwise, the encryption method used for the PPDL is not adequate since it doesn't allow the DL algorithm to learn from encrypted images.\\
In this paper, we have used MobileNetV2, which is a CNN composed of 53 layers. MobileNetV2 algorithm can be replaced by any other transfer learning algorithm. Figure \ref{fig:secureTraining} depicts the architecture of the secure training algorithm used in this study.

\section{Experimental Results}
This section is divided into three parts: dataset description, image encryption results, and CNN classification performances.
\subsection{Dataset Description}
In this paper, the COVID-19 Radiography database is used  \cite{chowdhury2020can,rahman2021exploring}. It is a public dataset that contains four classes, namely COVID-19, viral pneumonia, normal, and lung opacity. 
The dataset is randomly split into 80\% for training, 10\% for validation, and 10\% for testing. Table \ref{tab:dataset} shows the number of images used for training, validation, and testing for each class.

\begin{table}[h!]
\centering
  \caption{Dataset Partition}
  \label{tab:dataset}
  \begin{tabular}{lccc}
    \hline
    Class & Training& Validation& Testing\\
    \hline
   COVID-19 &2801 &350&350\\
          Lung Opacity &4801 &601&601\\
              Normal &5403 &675&676\\
    Viral pneumonia &963 &120&121\\

 \hline
\end{tabular}
\end{table}

\subsection{Implementation Details}

The experiments are carried out using a server with the following configuration properties: an x64-based processor, an Intel(R) Xeon(R) Gold 5218 CPU @ 2.30GHz, and a 512 GB RAM, with 8 GPUs (NVIDIA Quadro RTX 8000, 48 GB), running on Ubuntu 18.04 . The DL networks are programmed under Python 3.7 programming language. We used both the Keras 2.6 library and the TensorFlow-GPU 2.3 backend.

\subsection{Results}

\begin{figure}
  \centering
  \includegraphics[width=6cm]{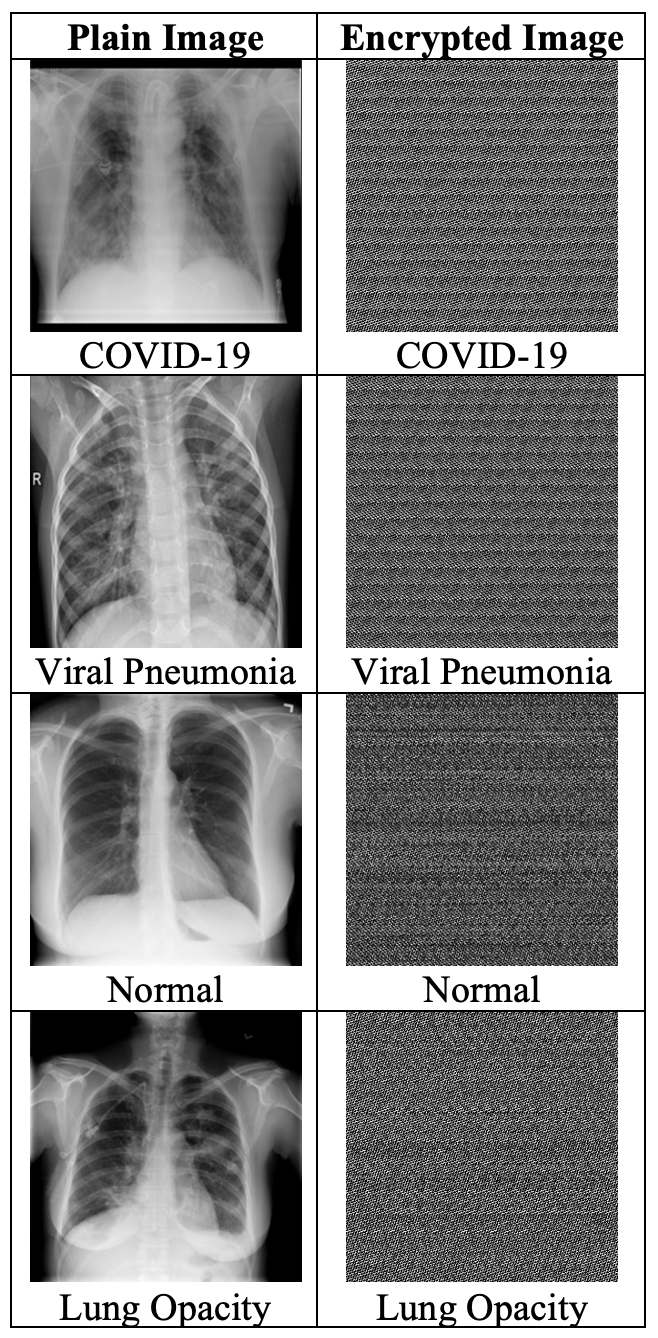}
  \caption{Sample of plain and encrypted images for each class.}
  \label{fig:sampleEncrypt}
\end{figure}

The first step is to encrypt the dataset using the Paillier encryption scheme. Figure \ref{fig:sampleEncrypt} depicts a sample of image encryption for the four classes COVID-19, viral pneumonia, normal, and lung opacity.

\begin{figure}
  \centering
  \includegraphics[width=8cm]{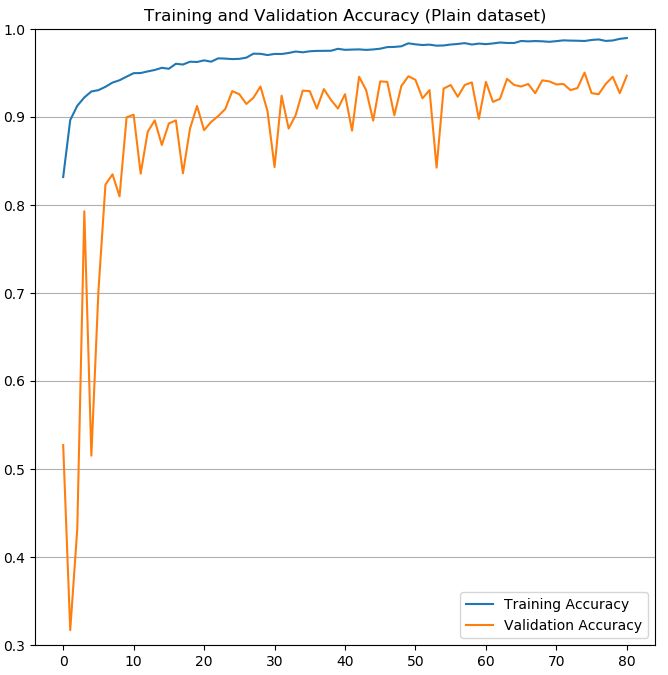}
  \caption{Training and validation accuracy of the plain dataset.}
  \label{fig:Acc_plain}
\end{figure}

\begin{figure}
  \centering
  \includegraphics[width=8cm]{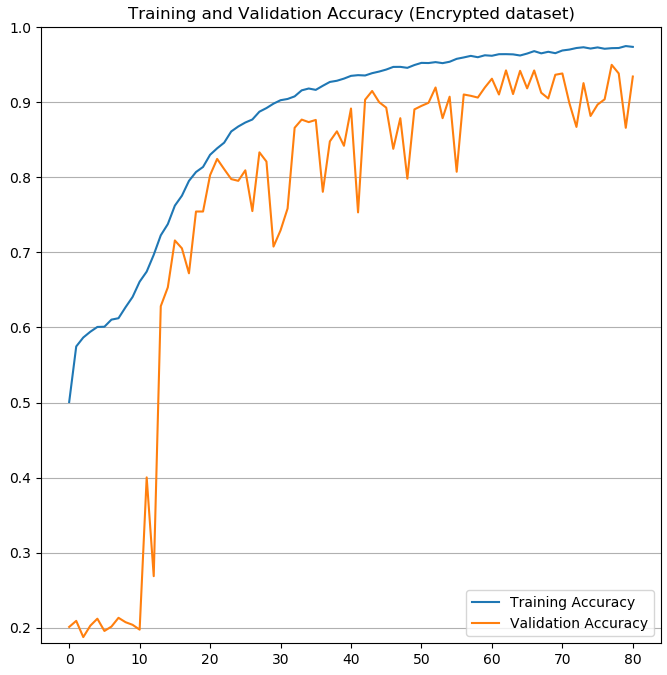}
  \caption{Training and validation accuracy of the encrypted dataset.}
  \label{fig:Acc_encrypted}
\end{figure}

\begin{figure}
  \centering
  \includegraphics[width=10cm]{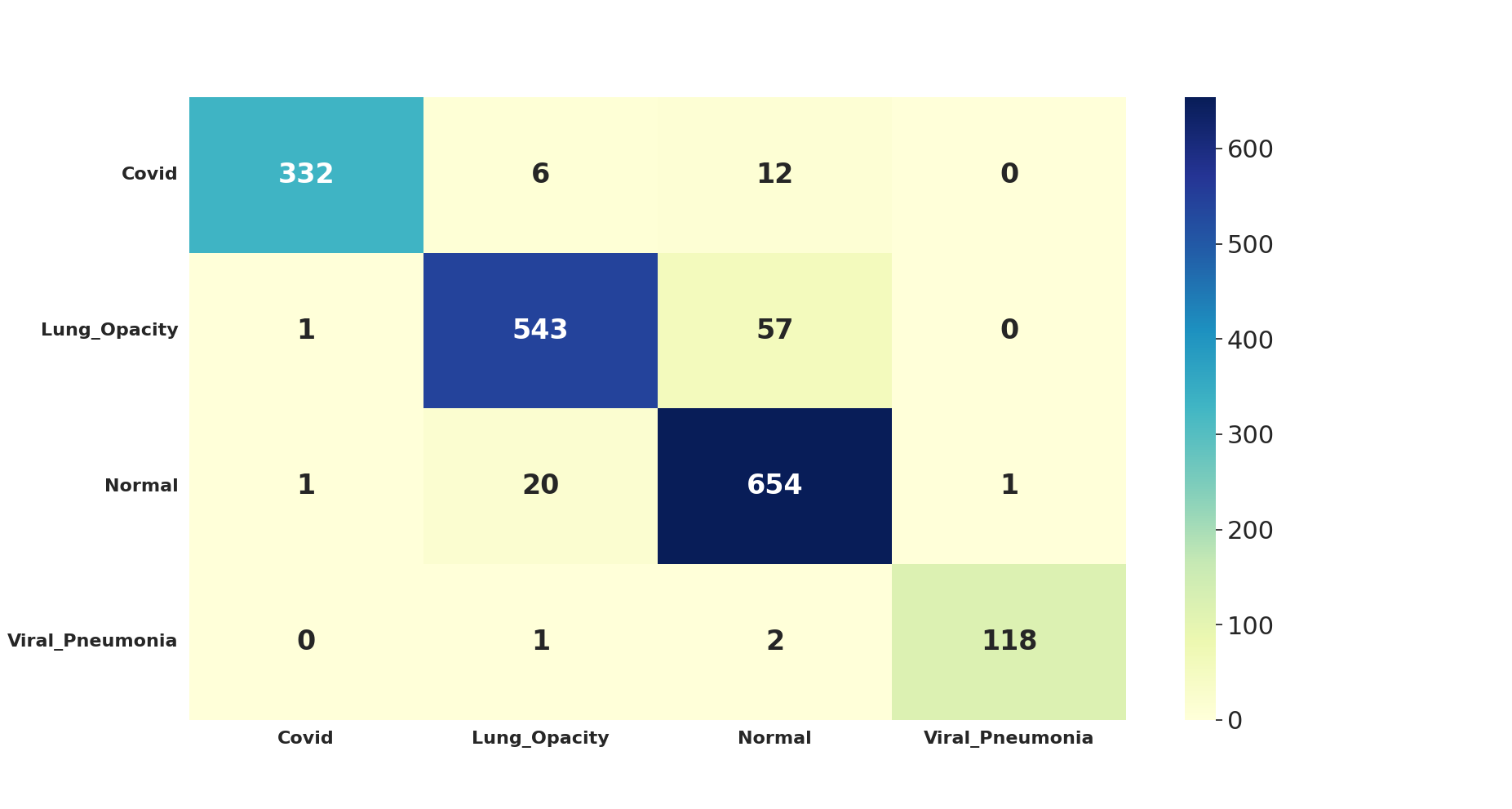}
  \caption{Confusion matrix on the test part of the dataset of plain images.}
  \label{fig:cm_plain}
\end{figure}

\begin{figure}
  \centering
  \includegraphics[width=10cm]{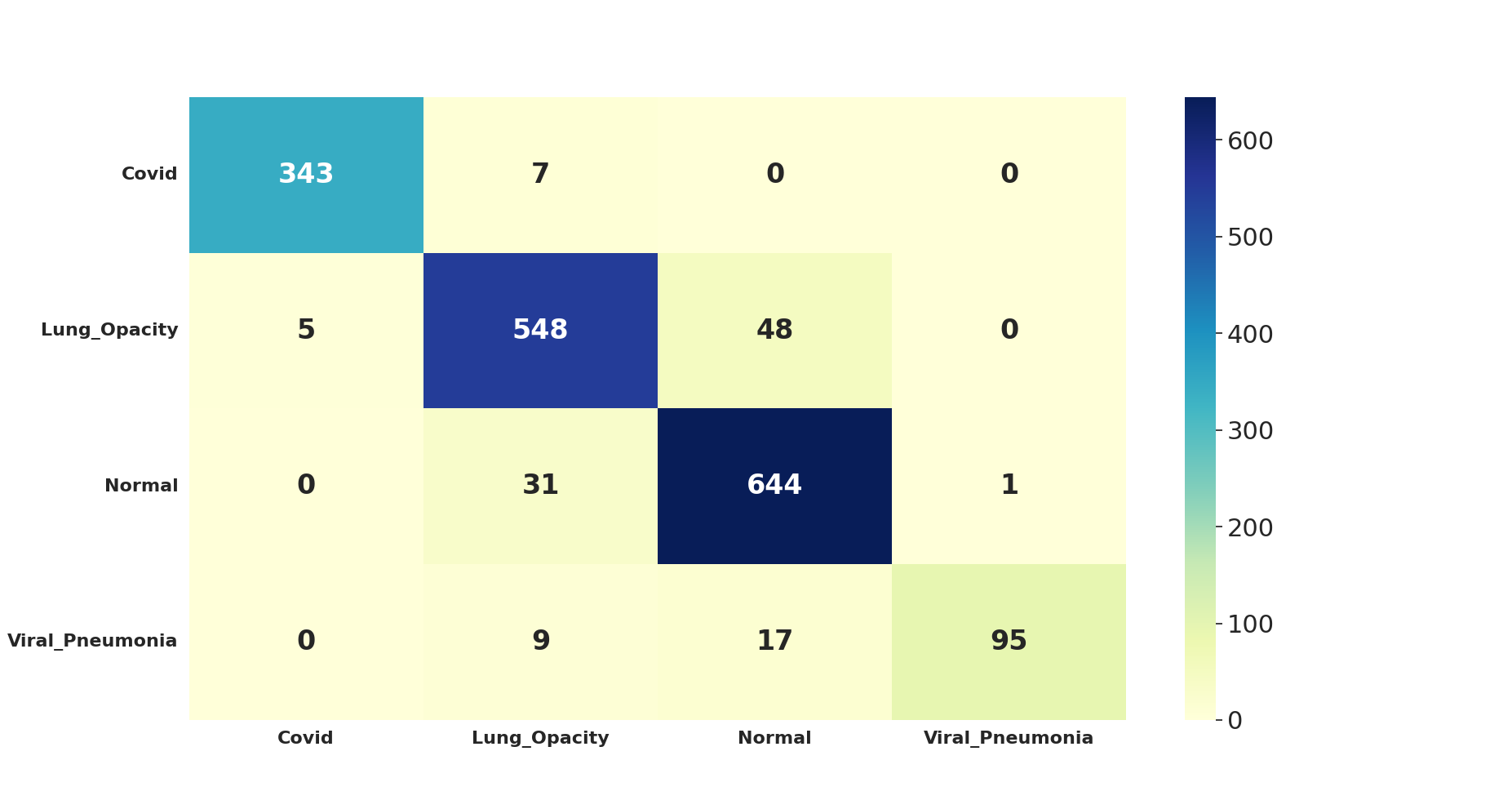}
  \caption{Confusion matrix on the test part of the dataset of encrypted images.}
  \label{fig:cm_encrypted}
\end{figure}

Then, the proposed DL architecture is applied to both plain and encrypted datasets. The training and validation accuracy of the MobileNetV2 with plain and encrypted data are depicted in Figures \ref{fig:Acc_plain} and \ref{fig:Acc_encrypted}, respectively. The maximum validation accuracy for the plain dataset is 95.04\%  obtained at the epoch 74. Whereas, the maximum validation accuracy for the encrypted dataset is 94.97\% obtained at the epoch 77.\\ 
In order to assess more robustly the performance on each dataset, we calculated the confusion matrix on the testing part which did not participate in any way in the training procedure nor in the selection of the best weights. Figures \ref{fig:cm_plain} and \ref{fig:cm_encrypted} show the resulting confusion matrices for the plain and encrypted datasets, respectively. In both cases, most misclassified images (76\% and 67\% of all misclassifications for the plain and encrypted datasets, respectively) are between the 'Normal' and 'Lung opacity' classes. 
Tables \ref{tab:results_plain} and \ref{tab:results_encrypted} show the detailed results, in terms of precision, recall, and f1-score, on the testing set of the plain and encrypted data, respectively. We notice that the 'Viral Pneumonia' class is the most affected by the encryption process, with an f1-sore decreasing from 98.3\% to 87.6\%, due to the reduced number of images for this class (7\% of the dataset) while the three other classes were much less affected.
The difference in terms of overall classification accuracy between encrypted (94.22\%) and plain (93.25\%) data is less than 1\% on the testing set. Hence, we can conclude that the Paillier encryption scheme ensures the privacy of data while also maintaining a highly accurate prediction of the class type in the COVID-19 Radiography dataset.

\begin{table}
\centering
\caption{Classification results on the testing set of the plain data.}
\label{tab:results_plain}
\begin{tabular}{lcccc}
\hline
                 & precision & recall & f1-score & support \\ \hline
Covid            & 0.994     & 0.949  & 0.971    & 350     \\ 
Lung\_Opacity    & 0.953     & 0.903  & 0.927    & 601     \\ 
Normal           & 0.902     & 0.967  & 0.934    & 676     \\ 
Viral\_Pneumonia & 0.992     & 0.975  & 0.983    & 121     \\ \hline
Accuracy         &           &        & 0.942    & 1748    \\ 
Macro avg        & 0.96      & 0.949  & 0.954    & 1748    \\ 
Weighted avg     & 0.944     & 0.942  & 0.942    & 1748    \\ \hline
\end{tabular}
\end{table}

\begin{table}
\centering
\caption{Classification results on the testing set of the encrypted data.}
\label{tab:results_encrypted}
\begin{tabular}{lcccc}
\hline
                 & precision & recall & f1-score & support \\ \hline
Covid            & 0.986     & 0.980  & 0.983    & 350     \\ 
Lung\_Opacity    & 0.921     & 0.912  & 0.916    & 601     \\ 
Normal           & 0.908     & 0.953  & 0.930    & 676     \\ 
Viral\_Pneumonia & 0.990     & 0.785  & 0.876    & 121     \\  \hline
Accuracy         &           &        & 0.932    & 1748    \\ 
Macro avg        & 0.951      & 0.907  & 0.926    & 1748    \\ 
Weighted avg     & 0.934     & 0.932  & 0.932    & 1748    \\ \hline
\end{tabular}
\end{table}



\section{Conclusion}
HE-based encryption has been used in many research works due to its significant privacy benefits. The increasing need of ensuring privacy-preserving of data while using DL techniques makes HE a very attractive topic to the research community. In this paper, we propose using the PHE-based Paillier algorithm to tackle the problem of the privacy of sensitive healthcare data when using DL algorithms. The Paillier encryption enables securing the data and preserving good classification accuracy. Experiments conducted on the COVID-19 Radiography database show an accuracy of 94.2\% for plain data and 93.3\% for encrypted data. In this sense, the proposed research can serve as a tool to classify encrypted data by non-trustworthy third parties without disclosing confidentiality. However, the main limitation of HE encryption is its slow computation, which remains the major problem of this technique. Further studies will focus on proposing hybrid encryption techniques such as FHE encryption and SSC that can be used in conjunction with one another.

\bibliography{bibdata}
\bibliographystyle{IEEEtran}
\end{document}